# Antisymmetric magnetoresistance and helical magnetic structure in compensated Gd/Co multilayer

Surendra Singh[1, 2,*], M. A. Basha[1, 2], C. L. Prajapat[2,3], Harsh Bhatt[1], Yogesh Kumar[1], M. Gupta[4], C. J. Kinane[5], J. Cooper[5], M. R. Gonal[6], S. Langridge[5] and S. Basu[1,2]

[1]Solid State Physics Division, Bhabha Atomic Research Centre, Mumbai 400085, India
[2]Homi Bhabha National Institute, Anushaktinagar, Mumbai 400094, India
[3]Technical Physics Division, Bhabha Atomic Research Centre, Mumbai 400085, India
[4]UGC DAE Consortium for Scientific Research, University Campus, Khandwa Road, Indore 452017, India
[5]ISIS-STFC, Rutherford Appleton Laboratory, Didcot OX11 0QX, United Kingdom
[6]Glass and Advanced Material Division, Bhabha Atomic Research Centre, Mumbai 400085, India

*surendra@barc.gov.in

Using spin dependent specular and off-specular polarized neutron reflectivity (PNR), we report the observation of a twisted helical magnetic structure with planar $2\pi$ domain wall (DW) and highly correlated magnetic domains in a Gd/Co multilayer. Specular PNR with polarization analysis reveals the formation of planar $2\pi$DWs below a compensation temperature ($T_{Comp}$), resulting to positive exchange bias in this system. Off-specular PNR with spin polarization showed development of magnetic inhomogenities (increase in magnetic roughness) for central part (thickness ~ 25-30 Å) of each Gd layer, where magnetization is aligned perpendicular (in-plane) to an applied field. These magnetic roughness are vertically correlated and results into Bragg sheet in spin flip channel of Off-specular PNR data, which is contributing towards an antisymmetric magnetoresistance at $T_{Comp}$ in the system. The growth and tunability of highly correlated magnetic inhomogeneities (roughness) and domain structure around $T_{Comp}$ in combination of twisted helical magnetic structure with planar $2\pi$DWs will be key for application in all-spin-based technology.

The current-induced manipulation of magnetic order through spin-orbit torque (SOT) has recently attracted great interest for the realization of magnetic memory and logic application devices with fast switching [1-5]. The Dzyaloshinskii-Moriya interaction [6], and the spin Hall effect via heavy-metal layers [7-9] were the major phenomena that attributed for large chiral spin torques. Exchange coupling torque (ECT) recently showed a significant enhancement of the spin-torque efficiency in artificial antiferromagnetic (AF) structures [10,11], which allows moving nanoscale magnetic domain walls (DW) with current at large velocities [10].

The compensated rare earth (RE)- transition metal (TM) alloys and heterostructures, where the RE and TM moments are aligned antiparallel due to the strong AF interaction and the total net moment tends to zero, are potential candidate materials for realizing devices with higher speed and density [12-16]. A class of ferrimagnets consisting of RE-TM alloys and heterostructures also have the potential to exhibit DW motion via an ECT [17,18]. Fast switching in compensated systems can further be influenced by magnetic [16,19] and optical [20] fields. Recently Vedmedenko et al. [21], have pointed out theoretically that nano-sized stable magnetic helices can be used for magnetic energy storage. Realization of magnetic helices with stable magnetic properties have also been studied theoretically [22] and experimentally [23] in exchange-coupled thin films and RE/TM multilayers, respectively. It is recognized that magnetization reversal and a magnetic helical configuration (planar $2\pi$DWs) in RE-TM multilayer with no external magnetic field around the compensation temperature ($T_{Comp}$, the temperature at which total moments of RE-TM multilayer tend to zero) is the key to manipulating magnetic devices [20, 23]. However the response of interface DWs in RE-TM heterostructures across the $T_{Comp}$ to magnetic fields and/or electric currents depends on the magnetic structure, magnetic phases, and domain evolution at the interfaces.

Here, we present strong evidence of helices in the form of planar $2\pi$DWs within both layers of Gd and Co in Gd/Co multilayers near the $T_{Comp}$ using polarized neutron reflectivity (PNR). PNR confirms an AF coupling between the Gd and Co layers. AF coupling along with the planar $2\pi$DWs formation below $T_{Comp}$ is responsible for the negative exchange bias observed in this multilayer. We also observed antisymmetric magnetoresistance (MR) at $T_{Comp}$, which is in contrast to earlier findings of similar effects in magnetic heterostructures with perpendicular magnetic anisotropy. Using spin dependent specular and off-specular PNR we demonstrate that antisymmetric MR at $T_{Comp}$ is a result of the evolution of highly correlated magnetic inhomogeneities (roughness) and magnetic domains of sub-micron length scale, in which the magnetization is aligned perpendicular (in-plane) to applied field.

The Gd/Co multilayer was grown using dc magnetron sputtering [24] on a Si (100) substrate (see Supplemental Material [25]) with a nominal structure: Si/[Gd(140Å)/Co(70Å)]$_{\times 8}$, where 8 is the number of repeats. Fig. 1 (a) shows the x-ray diffraction (XRD) patterns recorded for a Gd/Co multilayer along with single Co and Gd films. In contrast to earlier studies on Gd/Co multilayer systems [26,27], where a *hcp* structure for the Gd layer was observed, we found that the Gd layer has grown with a polycrystalline *fcc* structure [28]. However, Co has grown with a polycrystalline *hcp* structures. These results are consistent with an earlier study on Gd/Co multilayers grown on glass substrates [29]. Fig. 1(b) shows the x-ray reflectivity (XRR) data for the Gd/Co multilayer. Analysis of the XRR data provides the individual layer thickness, electron scattering length density (ESLD) and roughness at different interfaces of the multilayer [30,31]. Inset (i) of Fig. 1(b) shows the ESLD depth profile of the multilayer extracted from the XRR data. Parameters obtained from XRR are given in table S1 [25]. Small variations in roughness of each interface were considered to get the best fit to the XRR data. XRR results are corroborated by secondary-ion mass spectrometry (SIMS) measurements. SIMS data for a bilayer at the substrate interface of the multilayer is shown in the inset (ii) of Fig. 1(b), suggesting well defined layers.

Fig. 1(c) shows the in-plane magnetic hysteresis curves for the multilayer at different temperatures measured by a SQUID magnetometer. The observation of a very small coercive field ($H_c \approx 15$ Oe) at 300 K, where, only the Co is ferromagnetic, indicates the soft ferromagnetic nature of the multilayer. Magnetization data at different temperatures reveal a reduction in the saturation magnetization, an increase in $H_c$ and a shift of the hysteresis loop to negative magnetic field at low temperatures (Fig. 1(c) and Fig. S1 [25]). The shift of the hysteresis loop to negative field at low temperature reveals the negative exchange bias ($E_B$) in the multilayer. Fig. 1 (d) shows the variation of $H_c$ and $E_B$ with temperature. We observed the shift of the magnetic hysteresis loop below ~150 K ($E_B$ increases below this temperature), where $H_c$ starts decreasing.

The exchange bias at 5 K was further confirmed by measuring the in-plane magnetic hysteresis loops (Fig. 2(a)) of the multilayer after field cooling (FC) from room temperature in an applied magnetic field ($H$) of ~ ± 500 Oe. A shift of the hysteresis loop along the $H$ axis was observed towards negative (positive) fields on cooling the sample in a field of +500 Oe (-500 Oe), confirming the negative exchange bias in the system. Fig 2(b) shows the $M$(T) data from the multilayer under FC (both cooling (FCC) and warming (FCW) cycle) and zero field cooled (ZFC) conditions in an in-plane $H$ of 500 Oe, showing identical variation as a function of temperature. The *M*(T) data for the Gd/Co multilayer show a minimum in magnetization at a temperature around 125 K (~$T_{Comp}$).

Fig. 2(c - h) show the magnetoresistance (MR) data (%) [= $\frac{(\mathcal{R}(H)-\mathcal{R}(0))}{\mathcal{R}(0)} \times 100$, where, $\mathcal{R}(H)$ and $\mathcal{R}(0)$ are the resistance in the *H* and in zero field] at different temperatures in the longitudinal direction (*H* and current are in the same direction and along the plane of the film) as a function of the *H*. MR data measured on sweeping the *H* in the positive and negative direction are represented by blue (line with open triangles) and red (line with closed squares) curves, respectively. We observed different MR data as a function of temperature. The magnetic field dependence of the MR data at 300 and 200 K, show almost reversible (saturated) regions beyond the resistance peaks, similar to other magnetic multilayers. However the resistance peaks are at much higher field then coercive field (Fig. S2 [25]). We obtained irreversible and antisymmetric MR at 125 K. MR data at 100 K and below, again show the symmetric MR peaks. Although we observed additional irreversibility (separation) in the MR data when the *H* is scanned in opposite directions. Irreversibility in the MR beyond the peak region for different *H* direction decreases on decreasing the temperature.

In order to understand the correlation of the macroscopic magnetization (SQUID) and MR properties of the multilayer, we have studied the depth dependent structure and magnetization using PNR at different temperatures. The inset of Fig. 3 (a) shows the schematic of a PNR experiment with the ray diagram of scattering in *Q* space (Fig. S3 [25]). PNR measurements were carried out using the OFFSPEC reflectometer at the ISIS Neutron and Muon Source, RAL, UK. PNR data were taken in the *H* of +500 Oe at different temperatures upon warming the sample, after the sample was cooled at the same field from 300 to 5 K. Spin dependent specular ($Q_x = 0$) PNR with polarization analysis, i.e., non-spin-flip (NSF), $R^{++}$ and $R^{--}$, and spin-flip (SF), $R^{+-}$ and $R^{-+}$, reflectivities, are used to determine the magnitude and direction of the magnetization vector along the depth of the multilayer [32 - 34]. For these measurements NSF probes the projection of the magnetic induction vector parallel to the polarization direction, while SF are sensitive to the perpendicular component (Fig. 3(f)). Fig. 3(a)-(d) show the $R^{++}$ (●), $R^{--}$ (∆) and ($R^{+-} + R^{-+}$)/2.0 (*) reflectivities and corresponding fits (continuous lines) as a function of the wave-vector transfer $Q_Z$, normal to the sample surface, at different temperatures. The specular reflectivities data are collected up to a $Q_Z$ of ~0.08 Å$^{-1}$, which includes two Bragg peaks at $Q_Z$ ~ 0.03 Å$^{-1}$ (1$^{st}$ order) and 0.06 Å$^{-1}$ (2$^{nd}$ order) which corresponds to a bilayer periodicity of ~ 212 Å. Fig. 3(e) shows the nuclear scattering length density (NSLD) depth profile of the multilayer obtained from the specular PNR, which is consistent with the ESLD profile obtained from XRR data.

It is noted that Gd exhibits a large absorption for thermal neutron, which is wavelength dependent [36]. Using PNR data with (Fig. 3(a)) and without (PNR data up to larger $Q_Z$ ~0.16 Å$^{-1}$, Fig. S5 [25]) polarization analysis were used to fit the $\rho_N$ for Gd. We obtained $\rho_N$ = (1.05 + $i$. 3.42) ×10$^{-6}$ Å$^{-2}$ for Gd [25]. PNR data in Fig. S5 [25] also suggested an AF coupling of Gd-Co layer. The strong AF interaction in this system persists even for thicker Gd (~ 140 Å) and Co (~ 70 Å) layers, which may be due to the large exchange coupling ($J_{AF}$ = -2.1 × 10$^{-15}$ erg) [26] between Co and Gd spins as compared to the Zeeman energy ($\mu_B H$ = 4.6 × 10$^{-18}$ erg for $H$ = 0.5 kOe, field applied to the sample for the PNR measurements).

We didn't observe any SF (($R^{+-}$ + $R^{-+}$)/2.0) signal at 300 K (Fig. 3(a)), suggesting ferromagnetic Co with a magnetic scattering length density (MSLD) of ~ (3.55±0.16)×10$^{-6}$ Å$^{-2}$ (~ 1.52 µB/atom) and zero MSLD (moment) for the Gd layer. At 200 K we observed AF coupling between the Gd and Co layer, where the Co moments (MSLD ~3.78±0.17 ×10$^{-6}$ Å$^{-2}$ ~ 1.65 µB/atom) are aligned along the direction of the $H$ and the Gd moments (MSLD ~ -0.85±0.05×10$^{-6}$ Å$^{-2}$ ~ -1.40 µB/atom) are aligned antiparallel. The MSLD depth profiles at 300 and 200 K are shown in Fig. 3 (e). For comparison, negligible SF reflectivity is observed at 200 K. The solid line fit (Fig. 3(b)) for SF data at 200 K assumes a small inclination of the moments from the applied field by a small angle (~1-1.5 degree) suggesting the moments are essentially parallel to applied field at 200 K within error.

Strong SF signals are observed in the specular PNR data at 125 K and 5 K. Fig. 3(c) clearly suggest additional modulation in the PNR data at these low temperatures *e.g.*, a decrease in the intensity of $R^{++}$ data around the 1$^{st}$ order Bragg peak and a splitting of 2$^{nd}$ order Bragg peak for $R^{++}$ (for 125 K), suggesting a modification in the magnetic structure. Attempts to fit the PNR data at 125 and 5 K with homogeneous Gd and Co layers failed to reproduce the observed results and thus we considered a helical magnetic structure as depicted in Fig. 3(f). We have split the individual Co and Gd layer into sub layers with a constant magnetic moment within the Co and Gd layers but varying the angle of rotation of the magnetization with respect to the $H$ i.e. a helical structure. PNR data at 125 K reveal that the magnetization in both the Gd and Co layers rotate by 2π and form a planar 2π DW structure [23] as shown in Fig. 3(g). However, at the interfaces, Gd and Co are coupled antiferromagnetically, where Gd (Co) is aligned along (opposite) the $H$, which is consistent with the earlier findings for RE-TM system [35]. We have plotted the observed magnetization rotation angle in Fig. 3(g) for the sub layers within the Co and Gd layer, suggesting asymmetric rotation along the thickness of the Gd layer (i.e. the magnetization of the central part of the Gd layer is rotated by 90° instead of 180° as in the case of the Co layer). Therefore the depth dependent magnetic structure of the Gd/Co multilayer at $T_{Comp}$ exhibits the twisted helical structure. PNR measurements at 5 K suggested that the Co magnetization is still aligned opposite to the $H$ with a small variation in angle (180°±10°) for the Co sub layers. While the magnetization of the Gd sub-layer forms a $2\pi$ rotation within the Gd layer, which follows 0-$\pi$-0 rotation, instead of full $2\pi$ (0 to $2\pi$), as shown in Fig. 3(g).

While specular PNR as a function of $Q_Z$ provides depth profiles of the nuclear and magnetic structures, the lateral wave vector transfer $Q_X$ provides information on the correlation of lateral magnetic inhomogeneities (roughness and domains) in the sample plane, via off-specular scattering (see Supplemental Material [25]) [37-39]. Fig. 4(a) depicts the off-specular NSF, $R^{++}$ and $R^{+-}$ data ($Q_X$ - $Q_Z$ intensity map) at 5, 125 and 200 K. The $Q_X$ - $Q_Z$ intensity map for $R^{++}$ did not show any off-specular signals at different temperatures. However we obtained strong off-specular signals (Bragg sheet: intensity along $Q_X$ at Bragg positions) for SF ($R^{+-}$) mode at 125 K, which disappeared at high (200 K) as well as low (5 K) temperatures and hence suggesting a magnetic origin. Bragg sheets in the $R^{+-}$ reflectivity map at 125 K (= $T_{Comp}$), clearly indicate the development of magnetic inhomogeneities at interfaces that are vertically correlated. Variation of the off-specular intensity at the 2$^{nd}$ Bragg peak ($Q_Z$ ~ 0.06 Å$^{-1}$) as a function of $Q_X$ for different temperatures are also compared in Fig. 4 (c), justifying a magnetic source of the scattering at 125 K. Fig. 4(b) shows the corresponding simulated $R^{++}$ and $R^{+-}$ map at different temperatures. Simulation of the off-specular reflectivity have been performed using distorted wave Born approximation [25, 38,39]. Bragg sheets in the SF off-specular map at 125 K is well described by in-plane correlation length (magnetic domains) (~$\xi$) of 0.17 μm at the central part (thickness ~25-30 Å with magnetic roughness ~ 9 Å) of each Gd layer in the multilayer, for which the magnetic moment are aligned perpendicular (in-plane) to the $H$, as shown in Fig. 4(d) for a bilayer. We observed a fivefold increase in magnetic roughness ($\sigma_m$) for these interfaces at 125 K [25], as compared to that of 200 and 300 K. Moreover $\sigma_m$ for these intermediate Gd layer at 125 K is vertically correlated. We found smaller average lateral correlation length, $\xi$ (~0.01 μm) for all the interfaces below and above 125 K. Absence of Bragg sheet at 5 K indicate development of uncorrelated magnetic roughness.

AF-coupling of RE-TM systems has been attributed for the formation of planar DW (2$\pi$ DW) at the interfaces [40]. These 2$\pi$ DWs were responsible for the origin of exchange bias, $E_B$, in RE-TM multilayers [41, 42]. Our results for Gd/Co multilayer are consistent with these findings as we observed $E_B$ developing in the system near $T_{Comp}$, where there is a strong AF coupling between Gd and Co and specular PNR clearly suggested formation of magnetic helices with $2\pi$ DW within each Gd and Co layers. The $E_B$ increases at low temperature and we obtained the highest $E_B$ of ~ -75 Oe at 5 K. At 5 K, Co moments are mostly aligned opposite to the applied field and interface Gd moment are aligned opposite (aligned along $H$) to Co moments, while the Gd moments in Gd layer form a twisted helices with 0-$\pi$-0 configuration of the magnetization, therefore resulting in a maximum shift in hysteresis loop along negative field direction.

Another remarkable finding is the antisymmetric MR at $T_{Comp}$ and irreversibility in MR as a function of the $H$ around $T_{Comp}$. Different mechanisms are proposed to understand the MR effects in magnetic materials, however these effects share the common symmetry with respect to magnetization reversal, namely MR ($H$) = MR ($-H$). It is believed that the variation of multi-domain configuration during magnetization reversal process with MR ($H$) = -MR ($-H$) anomaly contributes to antisymmetric MR [43-45]. However there are mixed reports regarding the experimental conditions required for the observation of antisymmetric MR [43-45]. Cheng et al [43] observed the antisymmetric MR in Pt/Co multilayer structures and attributed it to specific configuration of mutually perpendicular direction of the domain wall, the current, and the magnetization. In contrast, Xiang et al [44] observe the antisymmetric MR only when the field and current were parallel to each other. It is noteworthy that we observed antisymmetric MR only at $T_{Comp}$ (125K) where we found highly correlated magnetic domains at the middle part of each Gd layer, using spin dependent off-specular PNR. We believe evolution of these highly correlated magnetic domains (with increase in magnetic roughness) at $T_{Comp}$, where magnetization are aligned perpendicular to the $H$, are responsible for antisymmetric MR. The antisymmetric MR can be explained qualitatively in line with ref. [43] as an increase in magnetic roughness (inhomogeneities) at $T_{Comp}$ which will perturb the current propagation (electric field) and the electric field will be reversed upon magnetization reversal. However the variation of helical magnetization as a function of temperature may contribute towards the additional irreversibility in MR across $T_{Comp}$.

In summary, we have observed negative exchange bias in Gd/Co multilayer below the compensation temperature ($T_{Comp}$ =125 K), which increases with a decrease in temperature. The exchange bias is due to the formation of planar domain walls across the thickness of the multilayer. Specular PNR provided detailed depth dependent magnetic structure of multilayer at different temperatures and suggested formation of planar $2\pi$DW, both within Co and Gd layer at $T_{Comp}$. PNR measurements also revealed formation of twisted helices across $T_{Comp}$ as a result of strong exchange coupling at the interfaces. Spin dependent off-specular PNR demonstrated evolution of magnetic inhomogeneities (increase in magnetic roughness) and magnetic domain of size 0.17 μm with magnetization direction perpendicular (but in the plane) to the $H$ in the central part of Gd layer at $T_{Comp}$, which are highly correlated along the thickness. These inhomogeneities and magnetic domain are responsible for antisymmetric longitudinal MR observed in Gd/Co multilayers at $T_{Comp}$. RE-TM multilayer as an artificial ferrimagnets can thus be a promising building block in devices with all-spin-based technology due to their helical magnetic structure and formation of planar 2πDW near compensation temperature.

## ACKNOWLEDGEMENT

Authors acknowledge the help of Nidhi Pandey and Prabhat Kumar of UGC-DAE-CSR Indore centre during deposition of the films. We thank the ISIS Neutron and Muon source for the provision of beam time (RB1768003 and OFFSPEC data: doi:10.5286/ISIS.E.92918790). Y. Kumar, would like to acknowledge Department of Science and Technology (DST), India for financial support via the DST INSPIRE Faculty research grant (DST/INSPIRE/04/2015/002938). C. L. Prajapat thanks the DST, India (SR/NM/Z-07/2015) for the financial support for performing experiment and Jawaharlal Nehru Centre for Advanced Scientific Research (JNCASR) for managing the project (SR/NM/Z-07/2015).

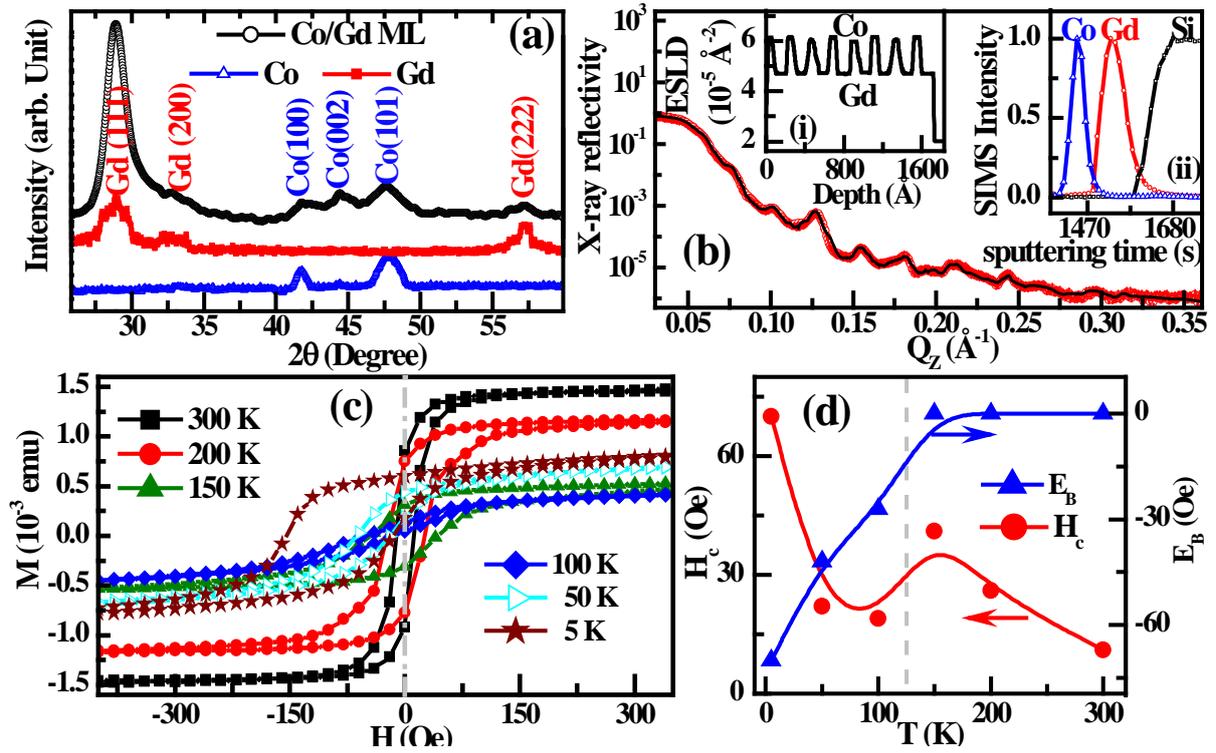

Fig. 1: (a) XRD scan for the Gd/Co multilayer, single Co and Gd films. (b) XRR data from the multilayer. Inset (i) and (ii) of (b) shows the electron scattering length density (ESLD) depth profile extracted from XRR data and SIMS data for a bilayer of the multilayer at substrate interface, respectively. (c) DC magnetization ($M(H)$) curve at different temperatures for the multilayer. (d) Variation of coercive field ($H_c$) and exchange bias ($E_B$) as a function of temperatures.

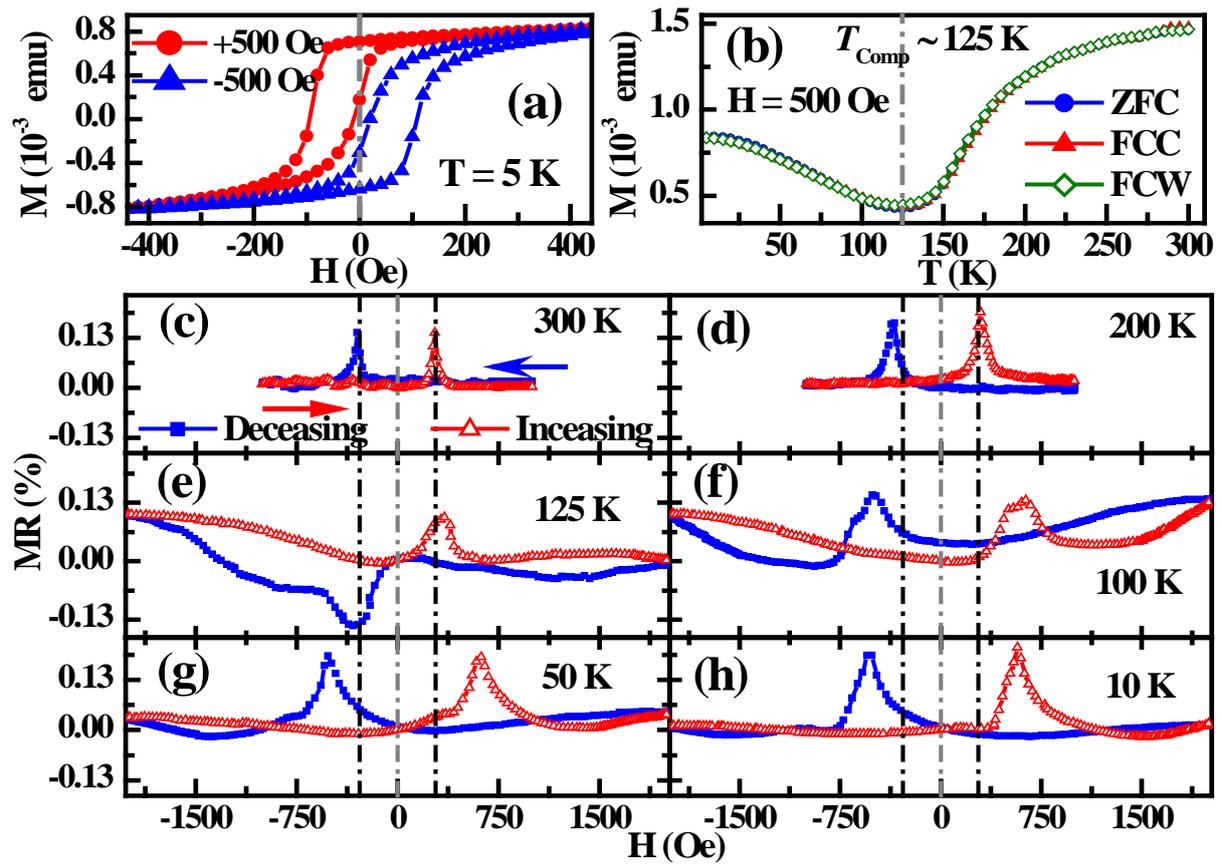

Fig. 2: (a) *M*(*H*) curves at 5 K on FC the sample at ±500 Oe. (b) Magnetization data as a function of temperature from multilayer for ZFC, FCC and FCW condition. (c-h) MR (%) data at different temperatures for the multilayer.

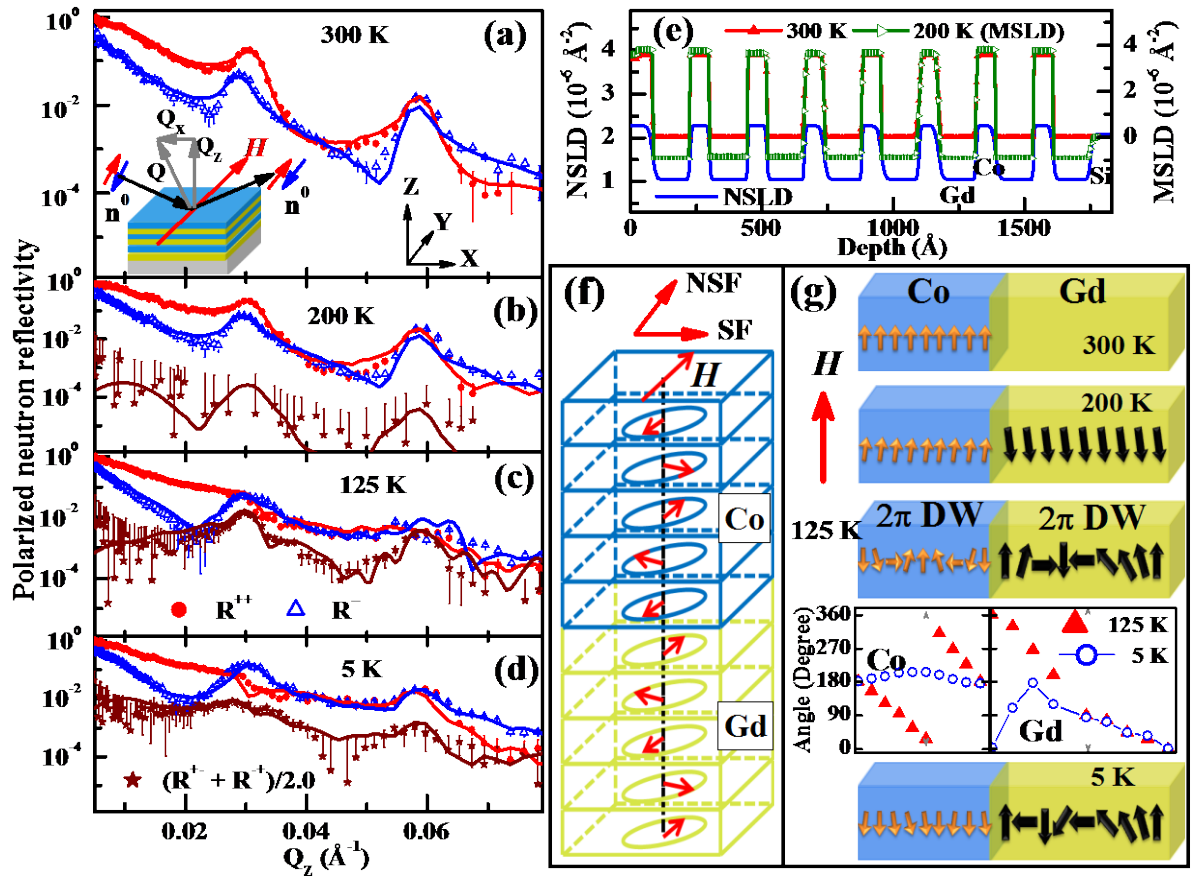

Fig. 3: (a-d) PNR data [non spin flip (NSF): R$^{++}$ (●), R$^{--}$ (Δ), and spin flip (SF): (R$^{+-}$ + R$^{-+}$)/2.0 (✱)] and corresponding fits (solid lines) from the Gd/Co multilayer at different temperatures. Inset of (a) shows the schematic of PNR experiment. (e) Nuclear and magnetic scattering length density (NSLD and MSLD) depth profiles of the multilayer. (f) schematic of helical magnetic structure. (g) Representation of magnetization in a bilayer of Gd/Co multilayer obtained from PNR data at different temperatures. The angle of rotation of the magnetization with respect to the $H$ in different sub layer within the Gd and Co layers at 125 and 5 K.

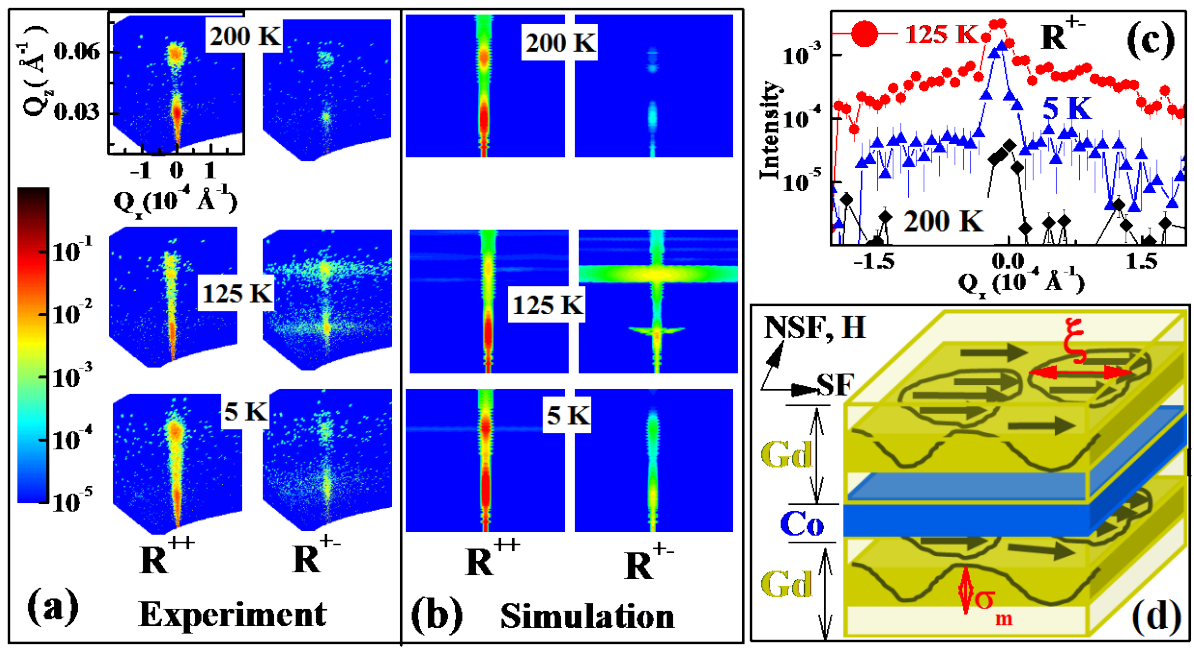

Fig. 4: (a) Off-specular PNR data ($Q_x$-$Q_z$ map) from the Gd/Co multilayer at different temperatures in non-spin flip (NSF), $R^{++}$ and spin-flip (SF), $R^{+-}$ modes. (b) Simulated profiles at different temperatures. (c) $R^{+-}$ intensity around second Bragg peak ($Q_z \sim 0.06$ Å$^{-1}$) at different temperatures. (d) Schematic of spin alignment of the Gd layer in a bilayer, contributing to the Bragg sheet in $R^{+-}$ intensity at 125 K.

# Antisymmetric magnetoresistance and helical magnetic structure in compensated Gd/Co multilayer

Surendra Singh[1,2,*], M. A. Basha[1,2], C. L. Prajapat[2,3], Harsh Bhatt[1], Yogesh Kumar[1], M. Gupta[4], C. J. Kinane[5], J. Cooper[5], M. R. Gonal[6], S. Langridge[5] and S. Basu[1,2]

[1]Solid Sate Physics Division, Bhabha Atomic Research Centre, Mumbai 400085, India
[2]Homi Bhabha National Institute, Anushaktinagar, Mumbai 400094, India
[3]Technical Physics Division, Bhabha Atomic Research Centre, Mumbai 400085, India
[4]UGC DAE Consortium for Scientific Research, University Campus, Khandwa Road, Indore 452017, India
[5]ISIS-STFC, Rutherford Appleton Laboratory, Didcot OX11 0QX, United Kingdom
[6]Glass and Advanced Material Division, Bhabha Atomic Research Centre, Mumbai 400085, India

*surendra@barc.gov.in

## *Sample Growth:*

The Gd/Co multilayer was grown using dc magnetron sputtering on a Si (100) substrate with a nominal structure: Si/[Gd(140Å)/Co(70Å)]$_{\times 8}$, where 8 is number of repeats. The multilayer was deposited under argon gas partial pressure of 0.2 Pa. Before deposition a base pressure of $1\times 10^{-5}$ Pa was achieved to avoid any contamination. The substrate was kept at room temperature during the growth of multilayer. For greater uniformity, substrate was rotated along its own axis at 60 rpm. Before deposition of Gd/Co multilayer we optimized the growth of the Gd and Co layers individually (single layer) on Si (100) substrate. The crystalline structure of the films grown on Si substrates was characterized by x-ray diffraction (XRD) technique. We obtained fcc crystalline structure for Gd and hcp crystalline structure for Co. The fcc crystalline structure observed for Gd in present Gd/Co multilayer might be one of the reasons for growth of well defined multilayer structure of Gd and Co, without formation of an alloy layer during deposition. However the growth parameters (argon partial pressure, substrate temperature etc.) will also influence the growth and layer structure of the heterostructures.

## Macroscopic magnetization (SQUID) measurements and magnetoresistance:

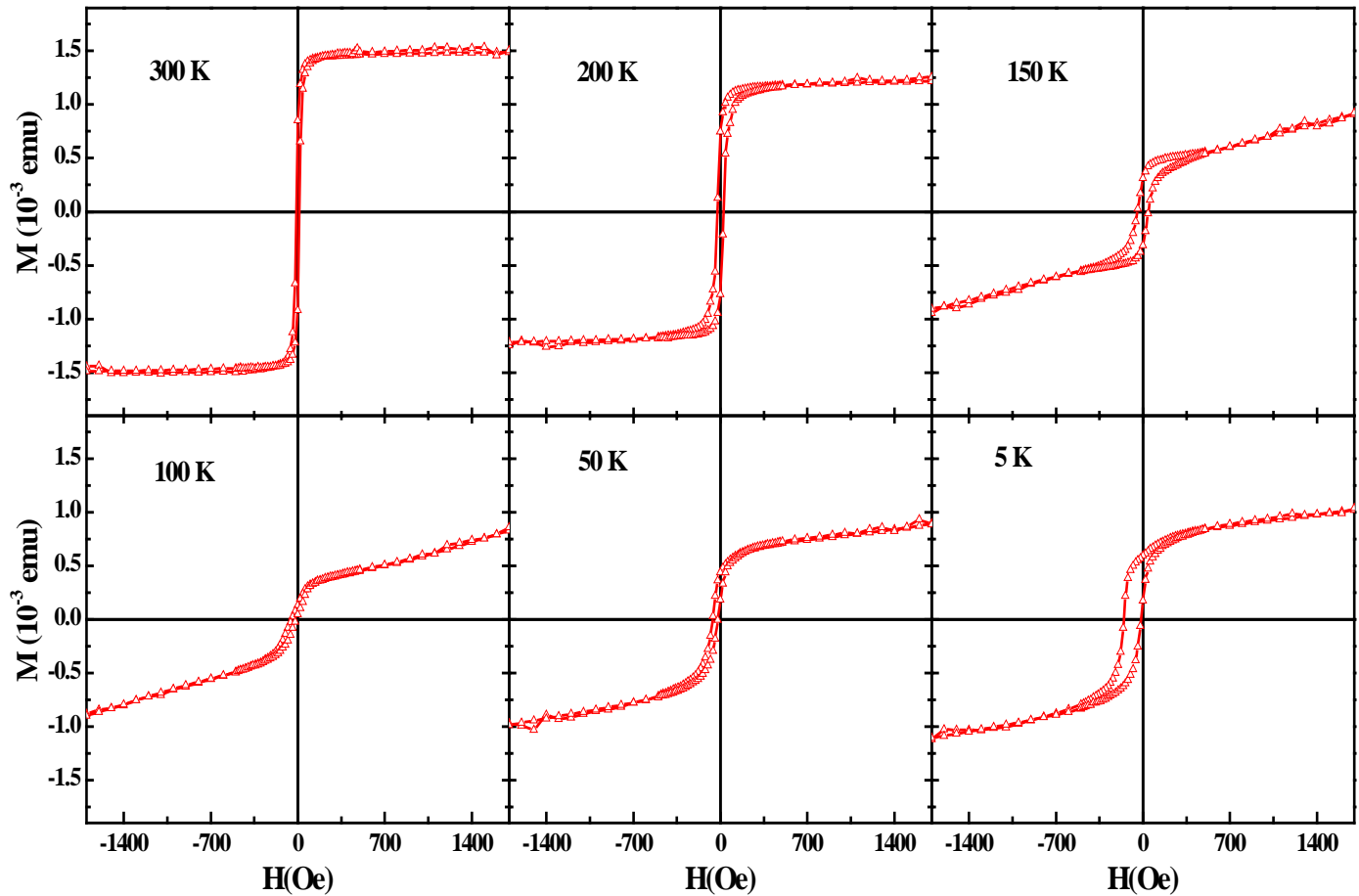

**Fig. S1: M(H) data at different temperatures from Gd/Co Multilayer.**

For magnetoresistance (MR) data we have measured the resistance using the four probe technique. MR measurements were carried out in longitudinal direction i.e. applied magnetic field and current are in the same direction and along the plane of the film. Fig. S2 shows the magnetization and MR at 300 K. It is evident from Fig. S2 that MR data at 300 K from multilayer exhibits resistance peaks at two magnetic fields ($H_p$), which is larger than the corresponding $H_c$. In general $H_p$ and $H_c$ for metallic multilayer coincide with each other. While the coercive field characterizes the random magnetization direction in the entire sample, the peak of the magnetoresistance curves is also an indication of the disordered magnetization configuration. The MR at different temperature also shows similar results where MR peak field ($H_p$) is larger than $H_c$. larger $H_p$ as compared to $H_c$ was also observed earlier in Co/Cu multilayer [1]. We also found that $H_p$ increases with decrease in temperature.

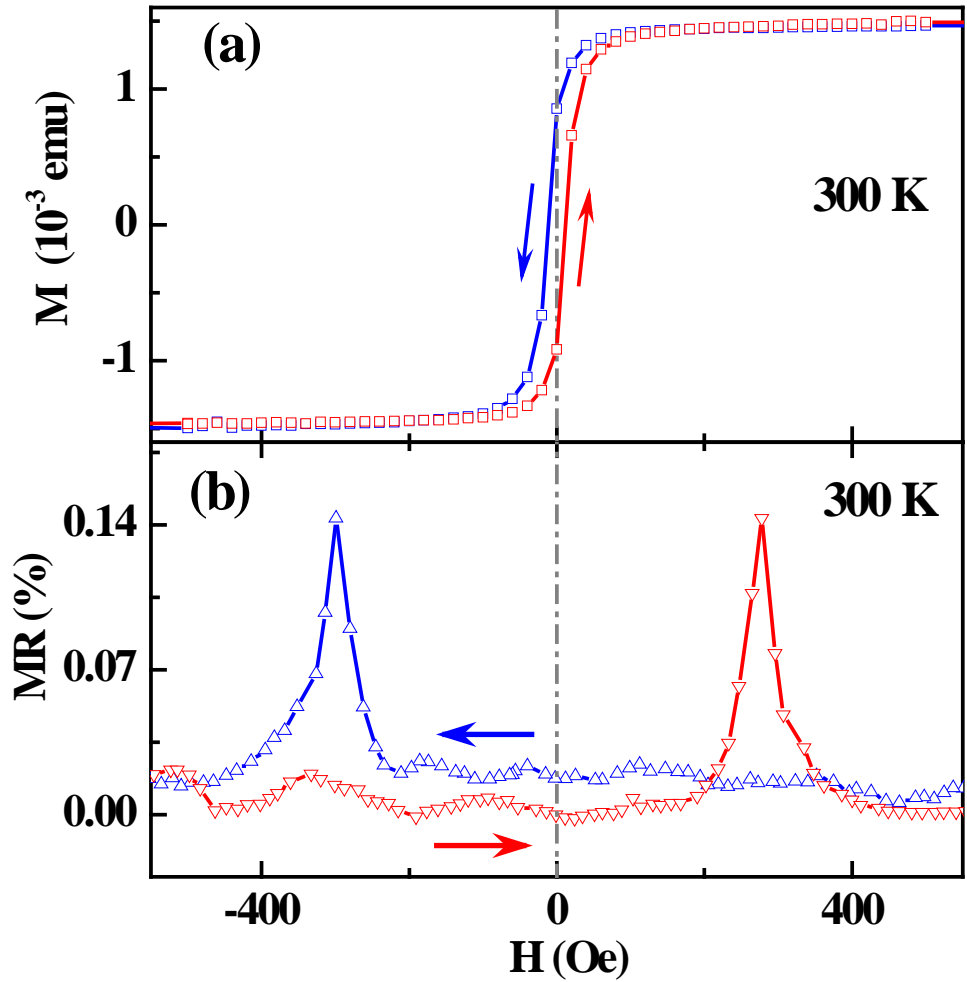

**Fig. S2:** Magnetization (a) and magnetoresistance (b) data as a function of magnetic field at 300 K.

### *X-ray and Polarized Neutron Reflectivity:*

X-ray reflectivity (XRR) and Polarized neutron reflectivity (PNR) are two nondestructive complementary techniques used to obtain the depth profiles of chemical and magnetic structure in multilayer sample with a depth resolution of sub nanometer length scale, averaged over the lateral dimension (~ 100 mm$^2$) of the sample [2-7]. Fig. S3 shows the schematic of reflectivity measurements. Reflectivity can be measured in two modes [7-12]: (a) specular reflectivity (where angle of incidence ($\theta_I$) is equal to angle of reflection ($\theta_F$), i.e.

$\theta_I = \theta_F$) and (b) off-specular reflectivity (where angle of incidence ($\theta_I$) is not equal to angle of reflection ($\theta_F$) i.e. $\theta_I \neq \theta_F$). However the plane of incidence and reflection are remaining same. The specular reflectivity is related to the square of the Fourier transform of the depth dependent (Z) scattering length density (SLD) profile $\rho(z)$ (normal to the film surface or along the Z-direction) [2-6]. For XRR, $\rho_x(z)$ is proportional to electron density and termed as electron scattering length density (ESLD) whereas for PNR without spin analysis of reflected beam, $\rho(z)$ consists of nuclear SLD (NSLD) and magnetic SLD (MSLD) such that $\rho^{\pm}(z) = \rho_n(z) \pm \rho_m(z) = \rho_n(z) \pm CM(z)$, where $C = 2.9109 \times 10^{-9}$ Å$^{-2}$ cm$^3$/emu, and $M(z)$ is the magnetization (emu/cm$^3$) depth profile [2-6]. $\rho_n(z)$ and $\rho_m(z)$ are NSLD and MSLD, respectively. The sign +(-) is determined by the condition when the neutron beam polarization is parallel (opposite) to the in-plane magnetization of the sample and corresponds to reflectivities, $R^{\pm}$.

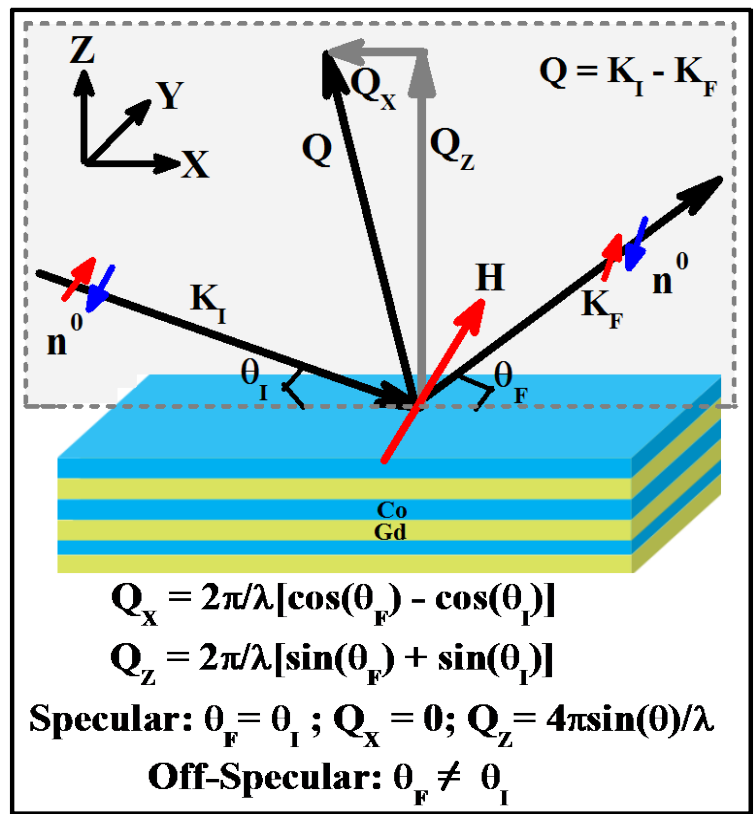

**Fig. S3**: Schematic of PNR experiment with a polarized neutron beam. Neutron beam incident (wave vector $K_I$) at the surface of the film with an angle of incidence of $\theta_I$ and reelected (wave vector $K_F$) at an angle of reflection of $\theta_F$. The difference between the incoming and outgoing wave vector is defined as momentum transfer vector $Q$ ( i.e. $Q = K_I - K_F$).

Specular XRR and PNR data can be fitted using a genetic algorithm based optimization program [13] which uses Parratt formalism [14]. Layers in a model consisted of regions with different electron SLD (ESLD). The parameters of a model included layer thickness, interface (or surface) roughness and ESLD. Errors reported for parameters obtained from XRR measurements represent the perturbation of a parameter that increased goodness of fit parameter corresponds to a 2σ error (95% confidence) [15].

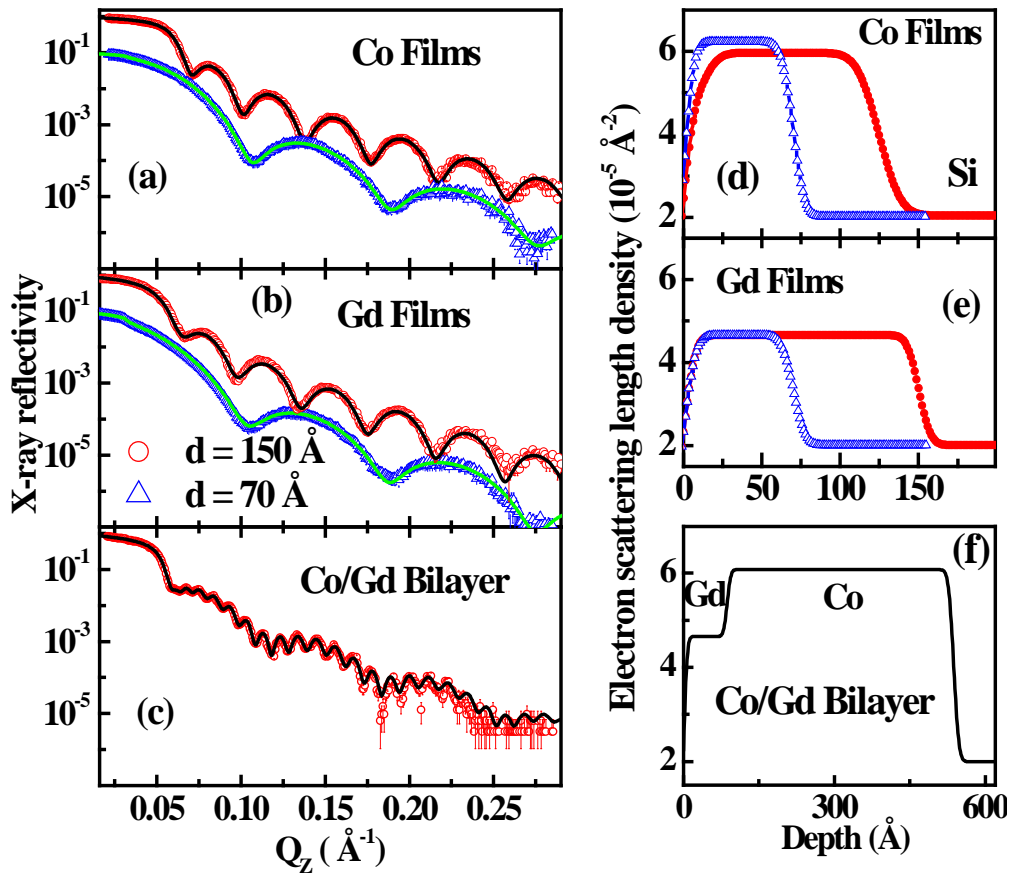

**Fig. S4**: (**a-c**) XRR data from single layer (Co, Gd), Co/Gd bilayer grown on Si substrates. (**d-f**) Electron scattering length density depth profiles for corresponding single layer and bilayer, which best fitted XRR data.

Fig. S4(a-c) show the XRR data from individual single layers of Co and Gd with different thicknesses (150 Å and 70 Å) as well as Gd (80 Å)/Co (450 Å) bilayer grown on Si substrate. Reflectivity is shifted for better visualization. Fig. S4(d-f) shows the corresponding electron scattering length density (ESLD) depth profiles, which best fitted XRR data shown in Fig. S4(a-c). These samples were grown to optimize the growth condition for Co, Gd and

Co/Gd bilayer on Si substrate. Also XRR was used to estimate the thickness of each layer in the heterostructure.

Layer structure parameters (thickness, ESLD and roughness) obtained from XRR data for Co/Gd multilayer are given in Table S1. We obtained an average roughness of ~ 8±2 Å and 4±1 Å for Gd/Co (Co on Gd) and Co/Gd (Gd on Co) interfaces. ESLD value for Co and Gd are very close to their bulk counterparts.

**Table S1: Parameters obtained from XRR and PNR measurements for Gd/Co multilayer grown on Si substrate.**

| Si(Substrate)/[Gd(143 Å )/Co(72 Å )]$_8$ multilayer | | | | | | |
|---|---|---|---|---|---|---|
| | From XRR | | | From PNR | | |
| layer | Thickness (Å) | Electron SLD ($10^{-5}$ Å$^{-2}$) | Averaged Roughness (Å) | Thickness (Å) | Neutron SLD ($10^{-6}$ Å$^{-2}$) | Averaged Roughness (Å) |
| Si (substrate) | - | 2.01±0.05 | 4.0±1.0 | - | 2.07±0.05 | 4.0±1.0 |
| Co | 72±3 | 6.10±0.08 | 4.0±1.0 | 73±3 | 2.25±0.04 | 4.0±1.0 |
| Gd | 143±4 | 4.60±0.10 | 8.0±2.0 | 144±3 | 1.04±0.04 | 8.0±2.0 |

## *PNR without polarization analysis:*

PNR measurements without spin analysis ($R^+$, spin up and $R^-$, spin down reflectivities) have been carried out at the neutron reflectometer POLREF at the ISIS, Rutherford Appleton Laboratory. Fig. S5 (a) shows the PNR, $R^+$ (●) and $R^-$ (Δ), data and corresponding fits (continuous lines) at different temperatures from Gd/Co multilayer. We have also collected temperature dependent PNR data without polarization analysis upto a larger $Q_Z$ (~0.16 Å$^{-1}$), while warming the sample from lowest temperature (~ 5K) after field cooling the sample in a field of 500 Oe from room temperature. It is evident from MSLD depth profile (Fig. S5(c-e)) that Gd and Co are antiferromagnetically coupled. At 300 K only Co is ferromagnetic. At 200K, Co moments are aligned along the field whereas Gd moments are aligned opposite to the applied field and vice versa at low temperatures (125 K and 10 K). It is noted that Gd exhibits large absorption ($\Sigma_{abs}$ ~ 65000 barn for neutron of wavelength = 2.5 Å) for thermal neutron, which is wavelength dependent [16]. Therefore we have used these data (upto larger Qz) to fit the nuclear coherent scattering length density, $\rho_N$, for Gd. We obtained $\rho_N$ = (1.05 + $i$. 3.42) ×10$^{-6}$ Å$^{-2}$ for Gd which along with $\rho_N$ for Co (Fig. S5 (b)) were kept constant while analyzing the PNR data at other temperatures and only magnetization was varied. The value of $\rho_N$ for Gd is very close to (0.96 + $i$. 3.12) ×10$^{-6}$ Å$^{-2}$ for a neutron of wavelength 2.6 Å, reported in ref. [16] and as reported by Lynn et.al. [16], it nearly remains constant for wavelength above ~2.6 Å.

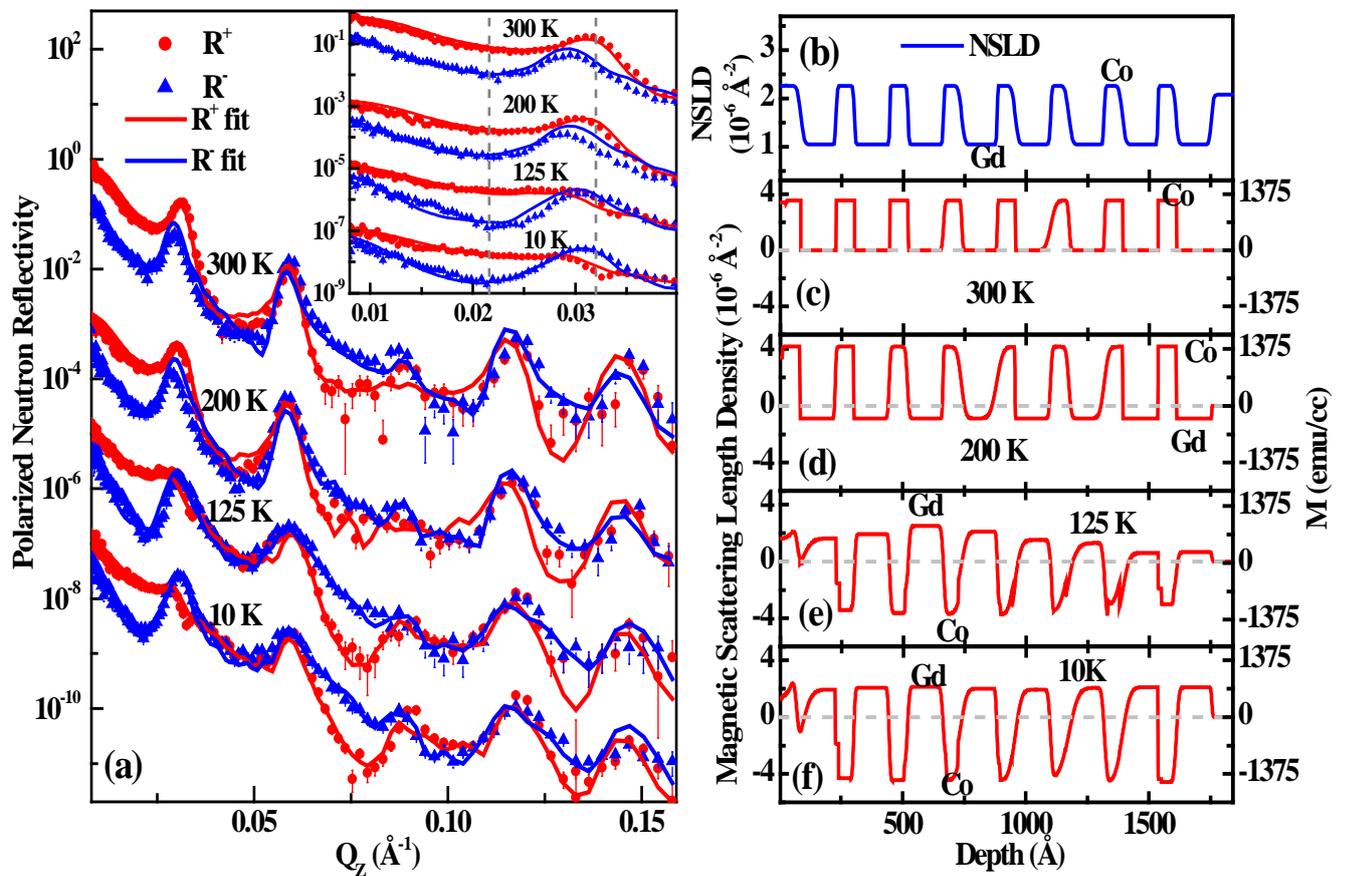

**Fig. S5**: (a) PNR data without polarization analysis from Gd/Co multilayer at different temperatures. Inset shows same data for a limited range of $Q_z$. (b) nuclear scattering length density (NSLD) depth profile. (c)-(f) magnetic scattering length density (MSLD) depth profiles at different temperatures.

*PNR with polarization analysis:*

PNR with spin analysis enables a quantitative, depth-resolved determination of magnitudes and directions of magnetization vectors in multilayer. Depth and laterally resolved magnetic models can be inferred from spin resolved specular and off-specular reflectivities via simulations of 2D reflectivity profiles as a function of temperature and external magnetic fields. Experimentally, a collimated, spin polarized neutron beam with a wave vector, $K_I = 2\pi/\lambda$, is impinging onto the sample surface at an incident angle $\theta_I$ (Fig. S3). Neutrons interact with the multilayer, and are reflected at an angle $\theta_F$ with a final moment $K_F$, which defines the momentum transfer $Q = K_I - K_F$. The neutron polarization vector is directed by the external field $H$ to lie within the plane of the sample perpendicular to the neutron propagation direction.

PNR measurements with spin analysis measurements were carried out the neutron reflectometer OFFSPEC, at the ISIS, Rutherford Appleton Laboratory. OFFSPEC allows simultaneous spin-dependent detection of specular and off-specular reflectivities. The instrument is equipped with a transmission-type supermirror polarizer, a pair of Mezei-type spin flippers, and a transmission-type supermirror analyzer for discriminating spin states of neutrons reflected into a broad range of wave vector transfers recorded over the 2D position sensitive detector (PSD).

The magnetic structure of multilayer manifests itself in the neutron reflectivity via differences in the spin dependent reflectivities, $R^{++}$, $R^{--}$, $R^{+-}$, and $R^{-+}$, where the first and second superscript denotes the direction of the incoming and reflected neutron polarization as parallel (+) or antiparallel (−) with respect to the external guide field ($H$) direction. The two nonspin-flip cross-sections, $R^{++}$ and $R^{--}$, are related to the magnetization components parallel to the applied (in-plane magnetization of sample). The remaining two spin-flip cross-sections, $R^{+-}$ and $R^{-+}$, are related to the magnetization components perpendicular to the applied field. Thus employing all spin dependent neutron reflectivity with polarization analysis provides the direction of in-plane magnetization along the depth of the heterostructures and interfaces. For depth dependent magnetic structure one measure specular reflectivities ($\theta_I = \theta_F = \theta$), which is defined as the conservation of in-plane momentum ($K_I = K_F$). Therefore the resultant momentum transfer ($Q$) is equivalent to momentum transfer component $Q_Z$ normal to the sample surface and is given as $Q = Q_z = \frac{2\pi}{\lambda}[\sin(\theta_f) + \sin(\theta_i)] = \frac{4\pi}{\lambda}\sin(\theta)$, where $\lambda$ is wavelength of the neutron. Off-specular reflectivity ($\theta_I \neq \theta_F$) originates from lateral structures, such as interfacial roughness and magnetic domains, which break the in-plane translational symmetry of the sample and lead to an in-plane momentum transfer $Q_X = \frac{2\pi}{\lambda}[\cos(\theta_f) - \cos(\theta_i)]$.

Specular PNR ($R^{++}$, $R^{--}$, $R^{+-}$, and $R^{-+}$) data as a function of temperature and magnetic field were analyzed with a genetic algorithm based optimization program [13] which uses a matrix [8] and supermatrix method [11]. Offspecular PNR data was analyzed using supermatrix method [9, 10, 12] within the framework of distorted wave Born approximation (DWBA). Using chemical structure (i.e. thickness, roughness at interfaces and number density) for multilayer as obtained from specular XRR, specular PNR data were fitted to obtain depth dependent lateral averaged structure (NSLD) and magnetization. Depth dependent magnetization structures as a function of temperature were obtained by fitting specular PNR data at different temperature by varying magnetization only while keeping the structure (NSLD) fixed. The structural parameters include thickness, real and imaginary part of NSLD and roughness. Magnetic structure is included via magnetic SLD (MSLD) depth profile and angle of rotation of magnetization (in the plane of the sample) along the thickness with respect to applied filed.

Off-specular reflectivity with a diffusion profile along $Q_X$ originates from structural/magnetic roughness and random in-plane magnetic domains at interfaces [9-12]. Like specular reflection, off-specular reflectivity is also a coherent phenomenon of constructive interference from neutron wave reflected from these in-plane inhomogeneities within the coherence volume of the sample. The off-specular spin dependent PNR reflectivity (intensity map or $Q_X - Q_Z$ map) can be calculated assuming a Gaussian-like lateral roughness structure factor, $S(Q_x)$ for vertically correlated interfaces [12]:

$$S(Q_x) = \frac{\sigma_m^2 \xi}{\sqrt{2\pi}} \exp\left(-0.5 \frac{Q_X^2}{\xi^2}\right),$$

Where $\sigma_m$ and $\xi$ are magnetic roughness and average lateral correlation length (magnetic domain size in lateral direction) at the interface. For uncorrelated vertical correlation of interfaces the structure factor is modified as:

$$S(Q_x) = \frac{\sigma_m^2 \xi}{\sqrt{2\pi}} \exp\left(-0.5 \frac{Q_X^2}{\xi^2}\right) \exp\left(-0.5 \frac{z - z'}{\xi_V}\right)$$

Where $z-z'$ is the depth and $\xi_v$ is vertical correlation length. For fully vertical correlation $\xi_v \sim$ infinity and exponential term containing vertical correlation length becomes 1.

To fit PNR data in off-specular mode we have used above formalism, where we have considered the in-plane magnetic domain distribution at interfaces, while keeping all the parameters obtained from specular PNR fixed. Splitting of each Gd and Co layers into sublayers make the system having similar chemical interfaces at all temperatures but different magnetic interfaces (each sublayer show different magnetic potential and hence can be defined with different set of off-specular parameters e.g. magnetic roughness, $\sigma_m$, $\xi$ and $\xi_v$) at low temperatures where we considered helical structure. Off-specular PNR measurements provide the lateral magnetic domain size via the average lateral correlation length, $\xi$. We have used this method to simulate the off-specular PNR data at different temperature and estimated the magnetic domain size ($\xi$) and magnetic roughness ($\sigma_m$) reported in the main paper. Fitting of off specular reflectivity suggested increase in magnetic roughness of central part (thickness ~ 25-30 Å) of each Gd layer where magnetization is rotated perpendicular to applied field. The in plane correlation length (~ domain size) for this part of Gd layer show a length scale of sub micron (~ 0.17 μm), which are correlated vertically.